\documentclass[conference]{IEEEtran}
\IEEEoverridecommandlockouts
\newcommand{\subparagraph}{}
\usepackage{braket,amsmath,graphicx,enumitem,bbm,cleveref,amsthm,titlesec,cite,dsfont}
\newtheorem{example}{Example}
\usepackage[left=0.8in, right=0.8in, top=0.8in, bottom=0.83in]{geometry}

\begin{document}

\title{Operating Characteristics for Binary Hypothesis Testing in Quantum Systems}

\author{\IEEEauthorblockN{Catherine Medlock (IEEE Student Member), Alan Oppenheim (IEEE Life Fellow), Isaac Chuang, Qi Ding\thanks{*This work was supported through the generosity of Analog Devices Inc., Texas Instruments Inc., and Bose Corporation.}}
\IEEEauthorblockA{Department of Electrical Engineering and Computer Science \\
Massachusetts Institute of Technology\\
77 Massachusetts Avenue, Cambridge MA, 02139}
}

%\vspace*{0.25in}
\maketitle

\begin{abstract}

Receiver operating characteristics (ROCs) are a well-established representation of the tradeoff between detection and false alarm probabilities in classical binary hypothesis testing. We use classical ROCs as motivation for two types of operating characteristics for binary hypothesis testing in quantum systems -- decision operating characteristics (QDOCs) and measurement operating characteristics (QMOCs). Both are described in the context of a framework we propose that encompasses the typical formulations of binary hypothesis testing in both the classical and quantum scenarios. We interpret Helstrom's well-known result \cite{helstrom1967detection} regarding discrimination between two quantum density operators with minimum probability of error in this framework. We also present a generalization of previous results \cite{eldar2002optimal, benedetto2008role} regarding the correspondence between classical Parseval frames and quantum measurements. The derivation naturally leads to a constructive procedure for generating many different measurements besides Helstrom's optimal measurement, some standard and others non-standard, that achieve minimum probability of error.

\end{abstract}

% no keywords

\section{Introduction} \label{sec:introduction}

Hypothesis testing in quantum systems is a well-studied problem in quantum information theory that plays an important role in quantum communication systems \cite{helstrom1967detection, chefles2000quantum, tej2018quantum, bae2015quantum, yuen1975optimum, weir2018optimal, helstrom1970quantum}. In that context each possible state represents a different transmitted message. More broadly, quantum hypothesis testing can be seen as a way to read out the information that has been computed by other quantum technologies and is contained in the state of a quantum system \cite{botelho2017quantum, spedalieri2016quantum}. Theoretical results regarding, for instance, strategies that minimize probability of error \cite{helstrom1967detection} or maximize the mutual information between input and output \cite{davies1978information}, are well-established.

Hypothesis testing in classical systems is also a problem that has been extensively studied \cite{helstrom1994elements,van2004detection}. Operating characteristics are often utilized to understand performance tradeoffs in classical binary hypothesis testing systems, and their properties are very well-understood. One of the most common types is the receiver operating characteristic (ROC), which displays the tradeoff between the probabilities of false alarm and detection in a classical binary hypothesis testing system. In contrast, operating characteristics of any kind are significantly less prevalent in the quantum binary hypothesis testing literature. In this paper we use classical ROCs as motivation for operating characteristics in the quantum setting.

Another line of thinking that is more common in the classical setting than in the quantum one is the notion of utilizing overcompleteness to create robustness to, for instance, noise and coefficient erasures. We briefly address the notion of overcompleteness in this paper in the context of known connections between classical frame theory and quantum measurement theory \cite{bodmann2017short, eldar2002optimal, eldar2001quantum, benedetto2008role, botelho2017solution, renes2004symmetric}. We then present an extension of previous results \cite{eldar2002optimal, benedetto2008role} regarding the correspondence between classical Parseval frames and quantum measurements.

\section{Outline} \label{sec:outline}

In Section \ref{sec:background} we review previous work on which this paper is based and introduce the necessary notation. In Section \ref{sec:operating_characteristics} we propose a hypothesis testing framework that encompasses the standard formulations of the binary hypothesis testing problems in both the classical and quantum scenarios. We then define two types of operating characteristics for the quantum scenario in the context of this framework. We refer to them as either quantum detector operating characteristics (QDOCs) or quantum measurement operating characteristics (QMOCs), depending on the parameter that is varied along the curve. QDOCs can be thought of as the direct analogue of classical ROCs. In Section \ref{sec:parseval_frames_and_povms} we summarize known connections between classical frame theory and quantum measurement theory, and derive a generalization of the results by Eldar and Forney \cite{eldar2002optimal} and Benedetto and Kebo \cite{benedetto2008role}. The generalization allows any quantum measurement to be represented in terms of many different classical Parseval frames. This naturally leads to a constructive procedure for generating many standard and non-standard measurements that achieve minimum probability of error when distinguishing between two fixed density operators. Concluding remarks are given in Section \ref{sec:conclusions}.

\section{Background} \label{sec:background}

In Section \ref{ssec:binary_hypothesis_testing} we point out a key difference between the typical formulations of binary hypothesis testing in the classical and quantum settings. The two postulates of quantum mechanics that are directly relevant to this paper are summarized in Section \ref{ssec:postulates}. In Section \ref{ssec:helstrom} we briefly review Helstrom's result \cite{helstrom1967detection} regarding discrimination between two quantum density operators with minimum probability of error.

\subsection{Typical Formulations of Binary Hypothesis Testing in Classical and Quantum Systems} \label{ssec:binary_hypothesis_testing}

In classical binary hypothesis testing the objective is to choose between two hypotheses given an observed value of a random variable sometimes referred to as the decision variable or score variable. Based on the value of the score variable, a final decision is made. We will denote the possible final decisions as `0,' indicating that one hypothesis was chosen, and `1,' indicating that the other was chosen.  Note that the decisions `0' and `1' should not be interpreted as the numerical values 0 and 1.

Classically, the method by which the score variable is generated can take many different forms depending on the application. For example, in a radar context a signal may be filtered, possibly by a matched filter, and then sampled in order to minimize the probability of error for signal detection in the presence of noise. In a clinical setting, the score variable may be generated by making a series of measurements on a patient and then combining them in an optimal way, such as by using a machine learning algorithm. In contrast, in a quantum mechanical setting the method of generating the score variable is often assumed to be a 2-outcome quantum measurement regardless of the application. As stated in Section \ref{ssec:postulates}, measurement of a quantum system can be described using positive operator-valued measures (POVMs), which are of course well-known in functional analysis \cite{berberian1966notes}. With this difference between the typical formulations of the classical and quantum binary hypothesis testing problems in mind, in Section \ref{sec:operating_characteristics} we propose a hypothesis testing framework that encompasses both.

\subsection{The Postulates of Quantum Mechanics} \label{ssec:postulates}

The fundamental differences between binary hypothesis testing in classical and quantum systems are embodied in the postulates of quantum mechanics. To be consistent with the literature we use Dirac's bra-ket notation, in which a vector $x$ is represented by the ket $\ket{x}$ and its Hermitian conjugate is represented by the bra $\bra{x}$. The two postulates of quantum mechanics as stated in \cite{nielsen2016quantum} that are particularly relevant to this paper are paraphrased below. The others are not directly relevant to our discussion here and are omitted.

\begin{enumerate}
    \item \emph{Quantum State Postulate:} The state of an isolated physical system can be described by a density operator $\rho$ that acts on a complex Hilbert space $\mathcal{H}$. We assume for convenience that $\mathcal{H}$ is finite dimensional with dimension $N$. Without loss of generality we may always associate $\mathcal{H}$ with $\mathds{C}^N$. $\rho$ is always a non-negative Hermitian operator that has trace equal to one. Thus, $\rho$ can be written in terms of its eigenbasis as
    \begin{equation}
        \rho = \sum_{i=0}^{N-1} \lambda_i \ket{w_i}\bra{w_i},
    \end{equation}
    where the $\{\ket{w_i}\}$ are orthonormal and the $\{\lambda_i\}$ are real, non-negative, and sum to 1. If $\rho$ has the form $\rho = \ket{w}\bra{w}$ for some vector $\ket{w}$, then $\rho$ is said to represent a pure state. Although the vector $\ket{w}$ is sometimes also referred to as a pure state, in this paper we will refer to it as a pure state vector to avoid confusion. If $\rho$ is not of the form $\rho = \ket{w}\bra{w}$, then $\rho$ is said to represent a mixed state.

    \item \emph{Quantum Measurement Postulate:} Quantum measurements are described by a collection $\{E_m\}$ of measurement elements that act on the state space $\mathcal{H}$ of the system being measured. We will assume for convenience that there are a finite number, $M$, of elements. Each element $E_m$, $0 \leq m \leq M-1$, corresponds to a different measurement outcome. Measurement elements are Hermitian, non-negative, and satisfy a completeness relation on $\mathcal{H}$,
    \begin{equation}
        \sum_{m=0}^{M-1} E_m = \mathds{1}_\mathcal{H},
    \end{equation}
     where $\mathds{1}_\mathcal{H}$ is the identity operator on $\mathcal{H}$. In functional analysis, a collection of operators that satisfies these three constraints is referred to as a positive operator-valued measure (POVM) \cite{berberian1966notes}.
    
    If the state of the system is described by the density operator $\rho = \sum_{i=0}^{N-1} \lambda_i \ket{w_i}\bra{w_i}$ immediately before a measurement with elements $\{E_m\}$, then with \mbox{probability}
    \begin{equation} \label{eq:pm}
        p(m) = \sum_{i=1}^N \lambda_i \bra{w_i}E_m\ket{w_i} = Tr[E_m \rho]
    \end{equation}
    the $m$th measurement outcome occurs. Here $Tr[\cdot]$ is the trace operator. The $m$th measurement outcome occurring indicates that the state of the system has collapsed to the $m$th post-measurement state, denoted by $\rho_m$. $\rho_m$ is determined by the pre-measurement state $\rho$ and the measurement element $E_m$. For more details, see Chapter 2 of \cite{nielsen2016quantum}.
\end{enumerate}

A POVM for which each $E_m$ has a one-dimensional range space is referred to as a rank-one POVM. A POVM for which each $E_m$ projects onto one of a complete set of orthogonal subspaces is referred to as a standard measurement, a projective measurement, or a von Neumann measurement. Each $E_m$ in a standard measurement is a projector onto its range, $R(E_m)$. This will be denoted as $E_m = \mathcal{P}_{R(E_m)}$. More generally, an operator that projects onto some subspace $\mathcal{V} \subset \mathcal{H}$ will be denoted by $\mathcal{P}_\mathcal{V}$.
\subsection{Binary Hypothesis Testing in Quantum Systems with Minimum Probability of Error} \label{ssec:helstrom}

This section summarizes Helstrom's well-known result \cite{helstrom1967detection} regarding discrimination between two fixed density operators with minimum probability of error. From this point forward, the word ``optimal'' will be used specifically to describe systems that achieve minimum probability of error unless otherwise specified. In the typical formulation of quantum binary hypothesis testing, we start with a quantum particle whose state is drawn from a source described by one of two known density operators depending on the value of the latent binary random variable $H$,
\begin{equation} \label{eq:quantum_input}
	\rho = \rho_0 \text{ if } H = H_0, \quad \rho = \rho_1 \text{ if } H = H_1.
\end{equation}
To distinguish between these two possibilities, a 2-outcome POVM with elements $\Pi_1$ and $\Pi_0 = \mathds{1}_\mathcal{H} - \Pi_1$ is performed. If the POVM is applied and the outcome associated with $\Pi_0$ occurs then the final decision is `0.' Otherwise the outcome associated with $\Pi_1$ occurs and the final decision is `1.' The relation between this formulation of binary hypothesis testing for quantum systems and our proposed hypothesis testing framework is given at the end of Section \ref{ssec:framework}.

The objective is to find the POVM elements $\Pi_1$ and $\Pi_0$ that minimize the probability of an error, which can be written as
\begin{equation} \label{eq:helstrom_prob_of_error}
    \mathds{P}(\text{error}) = \mathds{P}(H_1) - \mathds{P}(H_1) Tr\left[\Pi_1\left(\rho_1-C\rho_0\right)\right].
\end{equation}
where $C = \mathds{P}(H_0)/\mathds{P}(H_1)$ is the ratio of the prior probabilities for $H$. The result in \cite{helstrom1967detection} utilizes the orthonormal eigenvectors $\{\ket{z_m}, 0 \leq m \leq N-1\}$ and real eigenvalues $\{\eta_m, 0 \leq m \leq N-1\}$ of the operator $(\rho_1-C\rho_0)$. This operator or a scaled version of it is sometimes referred to as the Lagrange operator. It is shown in \cite{helstrom1967detection} that the probability of error is minimized when $\Pi_1$ and $\Pi_0$ are projectors onto the orthogonal subspaces $\mathcal{W}_1 = \text{span}\{\ket{z_m} : \eta_m \geq 0\}$ and $\mathcal{W}_0 = \text{span}\{\ket{z_m} : \eta_m < 0\}$,

\begin{equation} \label{eq:helstrom_subspace_criterion}
	\Pi_1 = \mathcal{P}_{\mathcal{W}_1}, \quad \Pi_0 = \mathcal{P}_{\mathcal{W}_0}.
\end{equation}
Note that those $\ket{z_m}$ for which $\eta_m = 0$ may be included in either $\mathcal{W}_0$ or $\mathcal{W}_1$ without changing the resulting probability of error. Since  $\mathcal{W}_0$ and  $\mathcal{W}_1$ are orthogonal, Helstrom's optimal measurement is by definition a standard measurement.

It is also noted in \cite{helstrom1967detection} that the optimal $\Pi_1$ and $\Pi_0$ can be written in terms of the $\{\ket{z_m}\}$ as
\begin{equation} \label{eq:helstrom_povm_zi}
	\Pi_1 = \sum_{m : \eta_m \geq 0} \ket{z_m}\bra{z_m}, \, \Pi_0 = \sum_{m: \eta_m < 0} \ket{z_m}\bra{z_m},
\end{equation}
which is a resolution of the projection operators in Equation \ref{eq:helstrom_subspace_criterion} in terms of orthogonal basis vectors. An equivalent way of achieving minimum probability of error is to use the $N$-outcome POVM whose elements are the terms in the sums,
\begin{equation} \label{eq:helstrom_n_outcome_povm}
E_m = \ket{z_m}\bra{z_m}, \quad 0 \leq m \leq N-1.
\end{equation}
 If the $m$th outcome occurs and $\eta_m > 0$, then the final decision is `1,' otherwise the final decision is `0.' If $\eta_m = 0$ then the final decision may be arbitrarily chosen as either `1' or `0.' By definition this is a rank-one, standard POVM.

\section[Operating Characteristics for Binary Hypothesis Testing in Quantum Systems]{Operating Characteristics for \\ Binary Hypothesis Testing in Quantum Systems} \label{sec:operating_characteristics}

In this section we use classical ROCs as motivation for operating characteristics for binary hypothesis testing in the quantum setting. First we propose a hypothesis testing framework in Section \ref{ssec:framework} that encompasses the typical classical and quantum scenarios as described in Section \ref{ssec:binary_hypothesis_testing}. The framework can easily be generalized to $m$-ary hypothesis testing for $m > 2$ by simply changing the number of possible final decisions. In Section \ref{ssec:classical_rocs} we describe classical ROCs in the proposed framework. We also summarize the result presented in \cite{medlock2019optimal} regarding the relationship between ROCs generated using what are referred to in \cite{medlock2019optimal} as score variable threshold tests (SVTs), and those generated using likelihood ratio tests (LRTs). QDOCs and QMOCs are discussed in Sections \ref{ssec:qdocs} and \ref{ssec:qmocs}, respectively. We emphasize that special cases of QMOCs, although not referred to in the same terminology, were presented by Bodor and Koniorczyk in \cite{bodor2017receiver}.

\subsection{Hypothesis Testing Framework for Classical and Quantum Systems} \label{ssec:framework}

\begin{figure}
    \centering
    \includegraphics[width=3.2in]{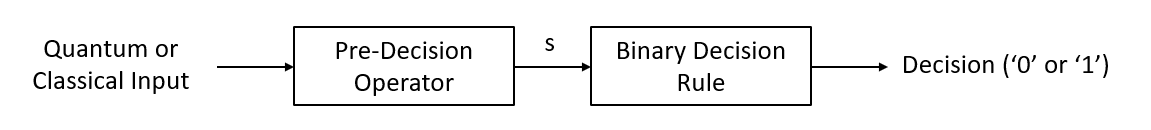}
    \caption{Framework for binary hypothesis testing in classical and quantum systems.}
    \label{fig:binary_hypothesis_testing_pipeline}
\end{figure}

The proposed framework is shown in Figure \ref{fig:binary_hypothesis_testing_pipeline}. The input is a classical or quantum object whose state depends on the latent binary random variable $H \in \{H_0, H_1\}$, with prior probabilities $\mathds{P}(H_0)$ and $\mathds{P}(H_1)$. As described in Section \ref{ssec:binary_hypothesis_testing}, in the quantum case the input is a quantum particle whose state is drawn from a source described by one of two known density operators, $\rho_0$ or $\rho_1$. Note that in future sections, we will use the phrase ``the input is $\rho$'' as a convenient shorthand for ``the input is a particle whose state is drawn from a source described by $\rho$.'' In the classical case, the input may be one of two known signals depending on whether $H = H_0$ or $H = H_1$, for example. It may also be a series of measurements on a patient whom we wish to test for an illness. The objective is to determine the value of $H$ in an optimal way with respect to some error criterion.

The first step in doing so is to obtain a sample of the score variable, denoted by $S$, using what we refer to as a pre-decision operator. The output of the pre-decision operator is a sample $s$ of $S$. In the classical case as described in Section \ref{ssec:binary_hypothesis_testing}, one example of a pre-decision operator is a matched filter followed by a sampler. In the quantum case, if the pre-decision operator is a POVM with elements $\{E_m, \, 0 \leq m \leq M-1\}$, then the score variable is discrete and its possible values correspond to the $M$ possible measurement outcomes. In order to assign concrete numerical values to the score variable, from this point forward we will arbitrarily associate the $m$th measurement outcome with the numerical value $m$, meaning that $s \in \{0, 1, \dots, M-1\}$. The conditional probability mass functions (PMFs) of $S$ will be denoted as
\begin{equation} \label{eq:p0_p1_defn}
	p_{S|H}(s | H=H_0) = p_0(s), \quad p_{S|H}(s | H=H_1) = p_1(s).
\end{equation}
A binary decision rule then uses the sample of $S$ to make a determination as to whether $H = H_0$, in which case the final decision is `0', or $H = H_1$, in which case the final decision is `1.' We again emphasize that the final decisions `0' and `1' should be interpreted as arbitrary labels and not as the numerical values 0 and 1. The design of the binary decision rule has of course been studied extensively in the context of classical decision theory. It can be viewed in terms of a partition of the sample space of $S$ into two decision regions, $\mathcal{D}$ and and its complement $\mathcal{D}^C$, where the output is
\begin{equation} \label{eq:D_defn}
	\text{`1' if } s \in \mathcal{D}, \quad \text{`0' if } s \in \mathcal{D}^C.
\end{equation}
We now connect this proposed framework to the typical formulation of the binary hypothesis testing problem as described in Section \ref{ssec:helstrom}. In the language of Figure \ref{fig:binary_hypothesis_testing_pipeline}, it is typically assumed that the pre-decision operator is a 2-outcome POVM with elements $\{E_0 = \Pi_0, \, E_1 = \Pi_1\}$ and that the subsequent binary decision rule has decision regions $\mathcal{D} = \{1\}$, $\mathcal{D}^C = \{0\}$. If the measurement outcome associated with $E_0$ occurs, then $s = 0$ and the final decision is `0.' Otherwise the measurement outcome associated with $E_1$ occurs, meaning that $s = 1$ and the final decision is `1.'

\subsection{Classical Operating Characteristics} \label{ssec:classical_rocs}

Classical ROCs are generated by fixing the pre-decision operator in Figure \ref{fig:binary_hypothesis_testing_pipeline}, varying the decision regions of the binary decision rule, and recording the resulting probabilities of false alarm ($P_F$) and detection ($P_D$). For fixed conditional PMFs $p_0(\cdot)$ and $p_1(\cdot)$ of the score variable and fixed decision regions $\mathcal{D}$ and $\mathcal{D}^C$, $P_F$ and $P_D$ are defined as
\begin{equation} \label{eq:pf_pd_defn}
	P_F = \sum_{s \in \mathcal{D}} p_0(s), \quad P_D = \sum_{s \in \mathcal{D}} p_1(s).
\end{equation}
One way of making a decision is to simply compare the score variable to a fixed threshold. We refer to this as a score variable threshold test (SVT) \cite{medlock2019optimal}. An SVT with threshold $\gamma$ decides `1' if $s \geq \gamma$ and `0' otherwise, i.e., its decision regions are
\begin{equation} \label{eq:svt_decision_region}
	\mathcal{D}_\text{SVT}(\gamma) = \{s : s \geq \gamma\}, \quad \mathcal{D}^C_\text{SVT}(\gamma) = \{s : s < \gamma\}.
\end{equation}
A different way to make a decision is to use a likelihood ratio test (LRT). An LRT with threshold $\eta \geq 0$ decides `1' if the likelihood ratio $p_1(s)/p_0(s) \geq \eta$ and `0' otherwise. The corresponding decision regions are therefore
\begin{subequations} \label{eq:lrt_decision_region}
\begin{alignat}{1}
	\mathcal{D}_\text{LRT}(\eta) &= \{s : p_1(s)/p_0(s) \geq \eta\} \\[7pt]
	\mathcal{D}^C_\text{LRT}(\eta) &= \{s : p_1(s)/p_0(s) < \eta\}.
\end{alignat}
\end{subequations}
It is well-known \cite{helstrom1994elements} that LRTs are optimal under the Neyman-Pearson criterion. That is, every LRT maximizes the probability of detection for a given upper bound on the probability of false alarm. It is also well-known \cite{helstrom1994elements} that every LRT minimizes probability of error for some combination of prior probabilities of the two hypotheses.

Even though SVTs are in general not equivalent to LRTs, SVTs are commonly used in many classical decision making settings. In \cite{medlock2019optimal} it was shown  (1) that concavity is a sufficient condition for the Neyman-Pearson optimality of an SVT ROC (that it is a necessary condition is widely known), and (2) that the LRT ROC can be reconstructed from a non-concave SVT ROC without knowledge of the conditional PMFs of the score variable. A constructive procedure was given for doing so. In the case where the conditional PMFs are known, or equivalently in the case where the SVT threshold of each point on the SVT ROC is known, the procedure implicitly provides an alternative way of constructing the Neyman-Pearson optimal ROC without directly applying LRTs: apply SVTs, then use the constructive procedure to generate the LRT ROC. In the case where the conditional PMFs are known, it also provides an alternative way of computing the LRT decision regions $\mathcal{D}_\text{LRT}(\eta)$ and $\mathcal{D}^C_\text{LRT}(\eta)$ for any $\eta \geq 0$, without requiring an analytical form for the likelihood ratio $p_1(s)/p_0(s)$.

In view of Figure \ref{fig:binary_hypothesis_testing_pipeline}, a performance curve analogous to classical ROCs can be made for the quantum case by fixing the POVM that constitutes the pre-decision operator and varying the decision regions of the binary decision rule. We refer to such an operating characteristic as a decision operating characteristic (QDOC). It is the direct quantum analogue of a classical ROC.

A different type of operating characteristic in the classical scenario is one generated by fixing the decision regions of the binary decision rule and varying the pre-decision operator. For example, consider the case where the input to the system is known to be one of two pre-determined signals in the presence of noise. Assume that both signals are bandlimited and that the pre-decision operator is a low-pass filter with cutoff frequency $\omega_c$ followed by a sampler. Further assume that the binary decision rule is an SVT with some fixed threshold $\gamma$. An operating characteristic can be generated by varying the cutoff frequency $\omega_c$ and plotting the corresponding probabilities of false alarm and detection. An analogous operating characteristic in the quantum case can be generated by keeping the decision regions of the binary decision rule fixed while varying the parameters of the POVM that constitute the pre-decision operator. We refer to this type of operating characteristic as a quantum measurement operating characteristic (QMOC).

In Sections \ref{ssec:qdocs} and \ref{ssec:qmocs}, we denote the eigendecompositions of $\rho_0$ and $\rho_1$ as
\begin{equation} \label{eq:rho0_rho1_defn}
	\rho_0 = \sum_{i=0}^{N-1} a_i \ket{x_i}\bra{x_i}, \quad \rho_1 = \sum_{i=0}^{N-1} b_i \ket{y_i}\bra{y_i},
\end{equation}
where the $\{\ket{x_i}\}$ and $\{\ket{y_i}\}$ each form orthonormal bases of $\mathcal{H}$ and the $\{a_i\}$ and $\{b_i\}$ satisfy ${0 \leq a_i \leq 1}$, $0 \leq b_i \leq 1$, $\sum_{i=0}^{N-1} a_i = 1$, $\sum_{i=0}^{N-1} b_i = 1$, as described in Section \ref{ssec:postulates}. It is sometimes useful to interpret the $\{a_i\}$ and the $\{b_i\}$ as two PMFs over the sets of vectors $\{\ket{x_i}\}$ and $\{\ket{y_i}\}$, respectively \cite{nielsen2016quantum}. For this reason, in what follows we will refer to the $\{a_i\}$ and $\{b_i\}$ as probabilities.

\subsection{Decision Operating Characteristics for Quantum Systems (QDOCs)} \label{ssec:qdocs}

A QDOC is generated by choosing a specific POVM for a given pair of density operators and varying the decision region of the binary decision rule. In Example \ref{ex:qdoc_1} below, the density operators $\rho_0$ and $\rho_1$ have the same eigenvectors but different eigenvalues. With a certain choice of POVM this implies that the conditional PMFs of the score variable correspond exactly to the eigenvalues of $\rho_0$ and $\rho_1$. As is well-known \cite{helstrom1967detection}, this is not the case if the eigenvectors of $\rho_0$ and $\rho_1$ are different.

\begin{example} \label{ex:qdoc_1} \normalfont
In this example we set $N = 8$ (recall that $N$ is the dimension of the Hilbert space) and $\ket{x_i} = \ket{y_i} = \ket{g_i}$, $0 \leq i \leq 7$, where $\ket{g_i}$ is the $i$th canonical basis vector for $\mathcal{H} = \mathds{C}^8$. The probabilities $\{a_i\}$ and $\{b_i\}$ are arbitrarily chosen to be the uniform distribution and an asymmetric triangular distribution, respectively, as shown in panel (a) of Figure \ref{fig:qdoc_ex1_final}. We have $a_i = 1/8$ for $0 \leq i \leq 7$ and $b_0 = 2/32, b_1 = 4/32, b_2 = 6/32, b_3 = 8/32, b_4 = 7/32, b_5 = 5/32, b_6 = 3/32, b_7 = 1/32$. We additionally assume that the pre-decision operator is an $8$-outcome standard POVM with elements $E_m = \ket{g_m}\bra{g_m}$, $0 \leq m \leq 7$. It can then be shown using Equation \ref{eq:pm} that the conditional PMFs of the score variable are $p_0(s) = a_s$ and $p_1(s) = b_s$, $0 \leq s \leq 7$. The LRT QDOC for this POVM is indicated by the solid black circles shown in panel (b) of Figure \ref{fig:qdoc_ex1_final}. The SVT QDOC is shown by the hollow black circles. Linear interpolation was used between the points to aid in visualization of the shapes of the curves. Of course, any operating point on any of the line segments could be achieved using randomization between two LRT or SVT decision regions \cite{helstrom1994elements}.

The SVT QDOC is clearly inferior to the LRT QDOC in Figure \ref{fig:qdoc_ex1_final}. However, the constructive procedure given in \cite{medlock2019optimal} can be used to reconstruct the LRT QDOC from the SVT QDOC without any knowledge of the PMFs $p_0(s)$ and $p_1(s)$. The same would be true for any two density operators $\rho_0$ and $\rho_1$. This leads to the interesting observation that for two arbitrary density operators $\rho_0$ and $\rho_1$ (whose eigenvectors may or may not be the same), if the eigenvectors $\{\ket{z_m}\}$ of the operator $(\rho_1-C\rho_0)$ are known but the eigenvalues $\{\eta_m\}$ are not, we may reconstruct Helstrom's orthogonal subspaces $\mathcal{W}_1$ and $\mathcal{W}_0$ in the following way: First choose an arbitrary ordering of the operators $E_m = \ket{z_m}\bra{z_m}$, $0 \leq m \leq N-1$. Generate the SVT QDOC associated with this ordering, then construct the LRT QDOC from it by following the procedure given in \cite{medlock2019optimal}. The resulting LRT QDOC will be the same regardless of the initial ordering. The procedure given in \cite{medlock2019optimal} provides a way of identifying the LRT decision regions associated with each point on the LRT QDOC. Identify the minimum probability of error operating point and its decision regions $\mathcal{D}$ and $\mathcal{D}^C$. According to Helstrom's result, $\mathcal{D}$ must contain all values of $m$ for which $\eta_m > 0$ and $\mathcal{D}^C$ must contain all values of $m$ for which $\eta_m < 0$. Values of $m$ for which $\eta_m = 0$ may be included in either $\mathcal{D}$ or $\mathcal{D}^C$. The subspace $\mathcal{W}_1$ is the span of the $\ket{z_m}$ for $m$ contained in $\mathcal{D}$, while the subspace $\mathcal{W}_0$ is the span of the $\ket{z_m}$ for $m$ contained in $\mathcal{D}^C$.

\begin{figure}
    \centering
    \includegraphics[width=3.2in]{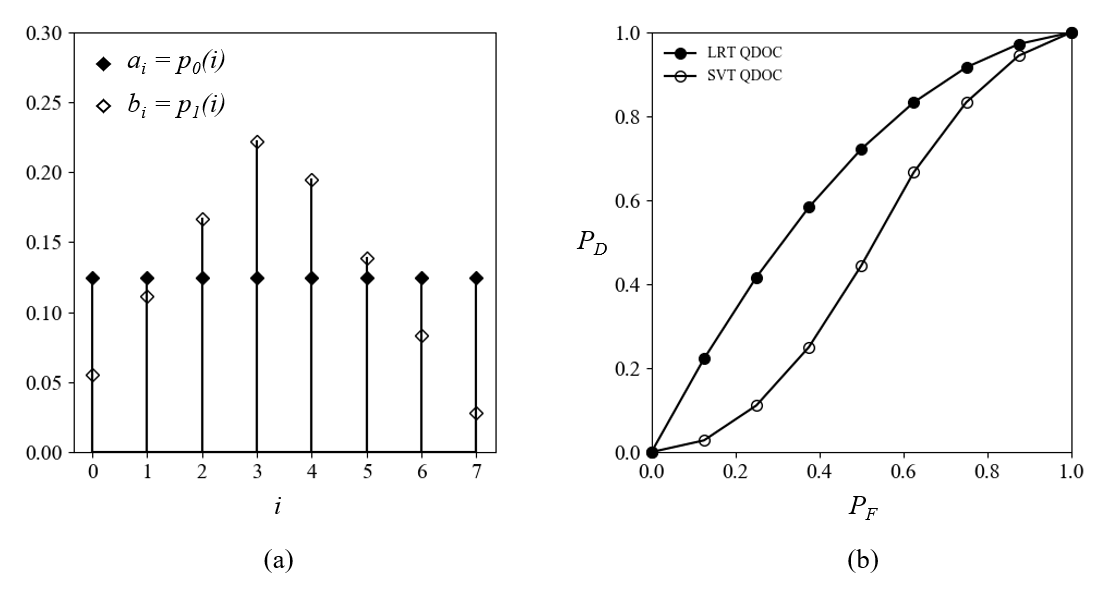}
    \caption{(a) Values of the $\{a_i\}$ and $\{b_i\}$ for two density operators $\rho_0$ and $\rho_1$, as defined in Equation \ref{eq:rho0_rho1_defn}. For this example the $\{a_i\}$ and $\{b_i\}$ correspond directly to the conditional PMFs of the score variable, $p_0(s) = a_s$ and $p_1(s) = b_s$. (b) QDOCs generated using LRT or SVT decision regions. Recall that $P_F$ and $P_D$ are defined in Equation \ref{eq:pf_pd_defn}.}
    \label{fig:qdoc_ex1_final}
\end{figure}
\end{example}

\subsection{Quantum Measurement Operating Characteristics (QMOCs)} \label{ssec:qmocs}

A QMOC is generated by fixing the decision regions of the binary decision rule and varying the POVM that constitutes the pre-decision operator. QMOCs in which the POVMs are chosen to be standard measurements were presented by Bodor and Koniorczyk in \cite{bodor2017receiver}. Note that while they are referred to in \cite{bodor2017receiver} as ``receiver operation characteristics,'' they are not the direct analogue of classical ROCs in our proposed framework. Example 2 below was inspired by one of the problems considered in \cite{bodor2017receiver}, in which $\rho_0$ and $\rho_1$ represent two mixed states of a qubit (i.e., $\mathcal{H}$ has dimension $N=2$) and standard measurements are used to distinguish between them. Unlike in \cite{bodor2017receiver}, we display the resulting curves for several values of the parameters of $\rho_0$ and $\rho_1$, in an attempt to suggest intuition about how these parameters affect the shape of the QMOC. This intuition is made concrete in Appendix \ref{sec:ellipse}, where we show that QMOCs generated in this way are ellipses. This fact was stated but not explicitly proven in \cite{bodor2017receiver}. We give formulas for the parameters of the ellipse in terms of the parameters of $\rho_0$ and $\rho_1$.

It is also interesting to consider generating a QMOC by plotting the minimum probability of error operating point for all possible sets of prior probabilities $\mathds{P}(H_0)$ and $\mathds{P}(H_1)$. In the case where $N=2$, it turns out that all such operating points lie on the ellipse generated using all possible standard measurements \cite{bodor2017receiver}. It was stated but not demonstrated in \cite{bodor2017receiver} that for $N>2$, the collection of optimal operating points for all possible sets of priors do not in general lie on an ellipse, but rather on a series of disjoint segments in the $P_F$-$P_D$ plane. We demonstrate this for a pair of arbitrarily chosen density operators with $N>2$ in Example 3.

\begin{example} \normalfont
We choose the $\{\ket{x_i}\}$ and $\{\ket{y_i}\}$ in Equations \ref{eq:rho0_rho1_defn} to be
\begin{equation} \label{eq:xi_yi_defn}
    \ket{x_0} = \begin{bmatrix} 1 \\ 0 \end{bmatrix}, \, \ket{x_1} = \begin{bmatrix} 0 \\ 1 \end{bmatrix}, \, \ket{y_0} = \begin{bmatrix} \cos(\alpha/2) \\ \sin(\alpha/2) \end{bmatrix}, \, \ket{y_1} = \begin{bmatrix} -\sin(\alpha/2) \\ \cos(\alpha/2) \end{bmatrix} 
\end{equation}
for some angle $\alpha$. We also arbitrarily set $a_0 = 1/15, a_1 = 14/15$. As in \cite{bodor2017receiver}, for fixed values of $\alpha$, $b_0$, and $b_1$, a 2-outcome standard POVM whose elements project onto the one-dimensional orthogonal subspaces spanned by the vectors
\begin{equation} \label{eq:qmoc_f1_f0}
	\ket{f_0} = \begin{bmatrix} -\sin(\theta/2) \\ \cos(\theta/2) \end{bmatrix}, \quad \ket{f_1} = \begin{bmatrix} \cos(\theta/2) \\ \sin(\theta/2) \end{bmatrix}
\end{equation}
is performed to distinguish between $\rho_0$ and $\rho_1$. In other words, the elements of the POVM that constitute the pre-decision operator are $E_0 = \ket{f_0}\bra{f_0}$ and $E_1 = \ket{f_1}\bra{f_1}$. If the outcome associated with $E_1$ occurs then the final decision is `1,' otherwise the outcome associated with $E_0$ occurs and the final decision is `0.' Thus, the decision regions of the binary decision rule are $\mathcal{D} = \{1\}$ and $\mathcal{D}^C = \{0\}$. Each QMOC is generated by varying the parameter $\theta$ in Equation \ref{eq:qmoc_f1_f0} from 0 to $2\pi$. According to Equation \ref{eq:pm}, the coordinates of the QMOC in terms of the angle $\theta$ are
\begin{subequations} \label{eq:qmoc_coordinates}
\begin{alignat}{1}
	P_F &= Tr[E_1 \rho_0] = a_0 \cos^2 (\theta/2) + a_1 \sin^2 (\theta/2) \\[7pt]
	P_D &= Tr[E_1 \rho_1] = b_0 \cos^2 ((\theta-\alpha)/2) + b_1 \sin^2 ((\theta-\alpha)/2).
\end{alignat}
\end{subequations}
In Appendix \ref{sec:ellipse} we show that Equations \ref{eq:qmoc_coordinates} is the parametric equation for an ellipse. The derivation also applies to certain cases where $N>2$. We now suggest intuition regarding the effect of the values of $\alpha$, $b_0$, and $b_1$ on the shape of the ellipse.

Figure \ref{fig:qmoc_ex2_final} shows a set of sample QMOCs generated using different combinations of $\alpha$, $b_0$, and $b_1$. In panel (a), $b_0$ and $b_1$ are arbitrarily fixed to $b_0 = 7/8, b_1 = 1/8$ while $\alpha$ is varied. Evidently, the value of $\alpha$ is an indication of the eccentricity of the ellipse: As $\alpha$ approaches $\pi$, the eccentricity of the ellipse increases. In panel (b), $\alpha$ is arbitrarily fixed to $\alpha = \pi/5$ while $b_0$ and $b_1$ are varied. The values of $b_0$ and $b_1$ apparently influence the vertical concentration of the ellipse: As $b_0$ and $b_1$ approach $1/2$, the ellipse becomes more concentrated around the line $P_D = 1/2$. Indeed, it is straightforward to show that the QMOC is inscribed in the rectangle with sides $P_F = min\{a_0,a_1\}$, $P_F = max\{a_0,a_1\}$, $P_D = min\{b_0,b_1\}$, $P_D = max\{b_0,b_1\}$.

\begin{figure}
    \centering
    \includegraphics[width=3.2in]{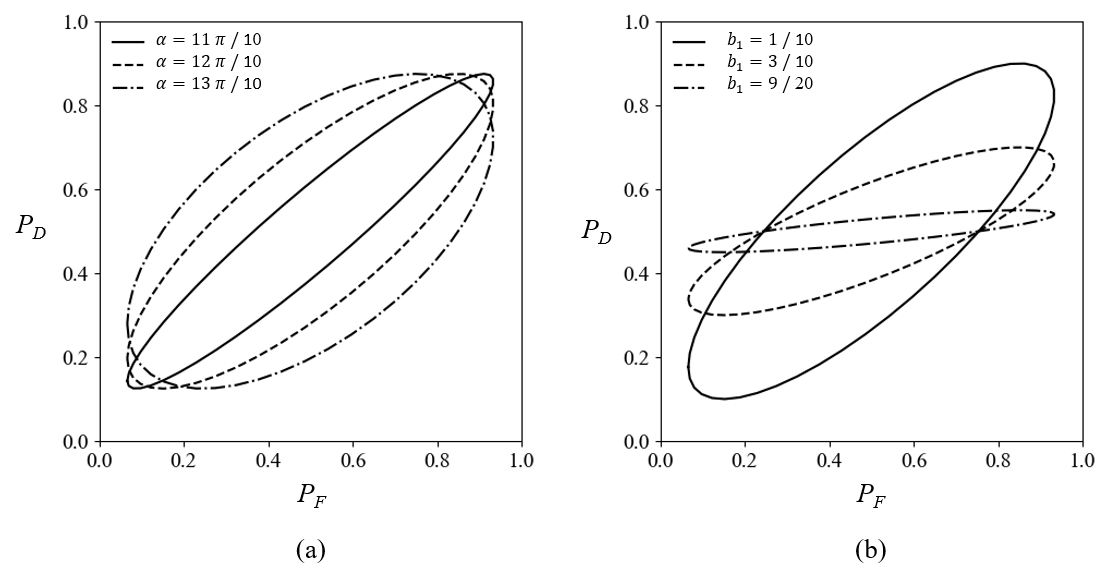}
    \caption{Sample QMOCs generated for two density operators with $N=2$ by varying the angle of a 2-outcome standard POVM. (a) QMOCs generated using different values of $\alpha$ with fixed values of $a_0, \, a_1, \, b_0,$ and $b_1$. (b) QMOCs generated using different values of $b_0$ and $b_1$ with fixed values of $\alpha, \, a_0$, and $a_1$.}
    \label{fig:qmoc_ex2_final}
\end{figure}
\end{example}

\begin{example} \normalfont
Here we use a pair of arbitrarily chosen density operators $\rho_0$ and $\rho_1$ with $N = 8$. We describe the collection of operating points that results from performing Helstrom's optimal measurement for a range of prior probabilities $\mathds{P}(H_1)$ and $\mathds{P}(H_0)$. For fixed priors, the optimal operating point has coordinates
\begin{equation}
    P_F = Tr[\Pi_1\rho_0], \quad P_D = Tr[\Pi_1\rho_1]
\end{equation}
where $\Pi_1$ is defined in Equation \ref{eq:helstrom_subspace_criterion}. The result is shown in Figure \ref{fig:all_standard_measurements}. The operating points form $(N-1)$ disjoint segments in the $P_F$-$P_D$ plane, in addition to the points $(0,0)$ and $(1,1)$. The value of $\mathds{P}(H_1)$ for which each operating point is optimal varies monotonically from left to right, with $(0,0)$ being optimal for $\mathds{P}(H_1) = 0$ and $(1,1)$ being optimal for $\mathds{P}(H_1) = 1$. The same properties apply to the collections of optimal operating points generated using other pairs of arbitrarily chosen density operators and other values of ${N>2}$.

Note that each pair of prior probabilities corresponds to a different decomposition of $\mathcal{H}$ in terms of Helstrom's orthogonal subspaces $\mathcal{W}_1$ and $\mathcal{W}_0$. The discontinuities between the segments in Figure \ref{fig:all_standard_measurements} correspond to changes in the dimension of $\mathcal{W}_1$ (equivalently, the number of non-negative eigenvalues of $(\rho_1-C\rho_0)$) \cite{bodor2017receiver}. The exception to this pattern is the case where $N>2$ and the eigenvectors of $\rho_0$ and $\rho_1$ with nonzero eigenvalues span a 1- or 2-dimensional subspace of $\mathcal{H}$. In that case the problem essentially reduces to the case where $N=2$, with the effective state space being the subspace spanned by the nonzero eigenvectors. In that situation as noted in \cite{bodor2017receiver} and derived in Appendix \ref{sec:ellipse}, the optimal operating points for all sets of priors lie on an ellipse.

\begin{figure}
    \centering
    \includegraphics[width=0.25\textwidth]{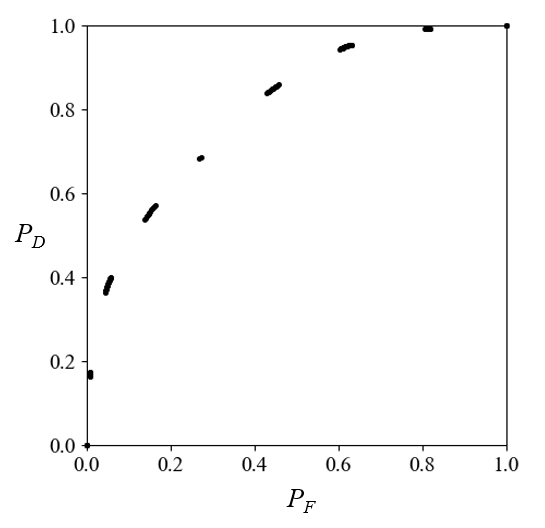}
    \caption{Minimum probability of error operating points for a pair of arbitrarily chosen density operators with $N=8$ for a range of prior probabilities, $0 \leq \mathds{P}(H_1) \leq 1$.}
    \label{fig:all_standard_measurements}
\end{figure}
\end{example}

\section{Parseval Frames and POVMs} \label{sec:parseval_frames_and_povms}

Helstrom's decision making strategy as well as the examples given in Sections \ref{ssec:qdocs} and \ref{ssec:qmocs} utilized standard measurements. Every standard measurement corresponds to a complete representation of $\mathcal{H}$ in terms of a set of orthonormal basis vectors or, more generally, a set of orthogonal subspaces. We may also wish to consider overcomplete representations of $\mathcal{H}$ in terms of sets of non-orthogonal frame vectors or non-orthogonal subspaces. It is for this reason that in many scenarios it is helpful to analyze quantum measurements using classical frame theory \cite{bodmann2017short, eldar2001quantum, eldar2002optimal, benedetto2008role, botelho2017solution, renes2004symmetric}. For the remainder of the paper we focus on a generalization of two results related to this area of research, as well as the interpretation of the generalization in our proposed hypothesis testing framework.

In Section \ref{ssec:eldar_forney_benedetto_kebo} we summarize the result by Eldar and Forney \cite{eldar2002optimal} regarding the expression of any rank-one POVM in terms of classical Parseval frame vectors. This was later generalized by Benedetto and Kebo \cite{benedetto2008role} to include non-rank-one POVMs. When interpreted in our proposed hypothesis testing framework, the result in \cite{benedetto2008role} naturally leads to the conclusion that given any hypothesis testing system whose pre-decision operator is an arbitrary POVM, it is always possible to construct an equivalent system whose pre-decision operator is a rank-one POVM. The word ``equivalent'' in this case means that both systems have the same probability of deciding `$i$' when the input is $\rho_j$, for $i,j \in \{0,1\}$. They then also have the same probability of error. We briefly discuss the notion of equivalent systems in more detail in Section \ref{ssec:equivalent_systems}. We then generalize the result in \cite{benedetto2008role} in Section \ref{ssec:standard_measurements} specifically for the case of standard measurements, since this yields valuable intuition for the general case. When applied to Helstrom's optimal 2-outcome standard measurement in Equation \ref{eq:helstrom_subspace_criterion}, the generalized result implies that there are many equivalent systems that can achieve minimum probability of error when distinguishing between two fixed density operators. The full generalization of the result in \cite{benedetto2008role} for potentially non-standard measurements is given in Section \ref{ssec:non_standard_measurements}.

\subsection{Previous Results Regarding Parseval Frames and POVMs} \label{ssec:eldar_forney_benedetto_kebo}

The result in \cite{eldar2002optimal} states that every rank-one POVM can be identified with a unique Parseval frame.\footnote{The converse is also shown to be true in \cite{eldar2002optimal}, but that is not relevant to our discussion here.} More specifically, the elements $\{E_m\}$ of any rank-one POVM can be written as
\begin{equation} \label{eq:rank_one_povm_parseval_frame}
E_m = \ket{f_m}\bra{f_m}, \quad 0 \leq m \leq M-1,
\end{equation}
where the collection of vectors $\{\ket{f_m}\}$ forms a Parseval frame for $\mathcal{H}$. Recall that a Parseval frame for a vector space $\mathcal{V}$ is a set of vectors $\{\ket{f_k}, 0 \leq k \leq K-1\}$ in $\mathcal{V}$ that satisfies the completeness relation $\sum_{k=0}^{K-1} \ket{f_k}\bra{f_k} = \mathds{1}_\mathcal{V}$, where $\mathds{1}_\mathcal{V}$ is the identity operator on $\mathcal{V}$. If $\mathcal{V}$ is a subspace of some larger vector space, then the frame vectors satisfy the equality $\sum_{k=0}^{K-1} \ket{f_k}\bra{f_k} = \mathcal{P}_\mathcal{V}$. Equation \ref{eq:helstrom_n_outcome_povm} is a special case of Equation \ref{eq:rank_one_povm_parseval_frame} where $M = N$, implying that the frame vectors $\{\ket{z_m}\}$ form an orthonormal basis of $\mathcal{H}$.

The more general result in \cite{benedetto2008role} states that the elements of an arbitrary (potentially non-rank-one) POVM can always be written as sums of outer products of the form $\ket{f}\bra{f}$. More specifically, it is pointed out in \cite{benedetto2008role} that the elements $\{E_m\}$ of an arbitrary POVM can be written in terms of their eigenvectors and eigenvalues as
\begin{equation} \label{eq:benedetto_and_kebo}
E_m = \sum_{k=0}^{N-1} \lambda_{mk} \ket{v_{mk}} \bra{v_{mk}} = \sum_{k=0}^{N-1} \ket{\tilde{v}_{mk}} \bra{\tilde{v}_{mk}},
\end{equation}
where $\ket{\tilde{v}_{mk}} = \sqrt{\lambda_{mk}}\ket{v_{mk}}$. Since $\sum_{m=0}^{M-1} E_m = \mathds{1}_\mathcal{H}$, this implies that the collection of vectors $\{\ket{\tilde{v}_{mk}}\}$ forms a Parseval frame for $\mathcal{H}$. Equation \ref{eq:helstrom_povm_zi} is a special case of Equation \ref{eq:benedetto_and_kebo} where the $\{E_m\}$ form a standard measurement, implying that each $\lambda_{mk}$ is either 0 or 1.

In Section \ref{ssec:equivalent_systems} we consider the case where the POVM $\{E_m\}$ is the pre-decision operator of a binary hypothesis testing system. We describe how the frame vectors $\{\ket{\tilde{v}_{mk}}\}$ associated with that POVM can be used to construct an equivalent system whose pre-decision operator is a rank-one POVM. In Sections \ref{ssec:standard_measurements} and \ref{ssec:non_standard_measurements}, we show that the Parseval frame associated with a particular POVM is not unique.

\subsection{Equivalent Systems in Our Proposed Hypothesis Testing Framework} \label{ssec:equivalent_systems}

We start by showing that, given a binary hypothesis testing system consisting of a 2-outcome POVM followed by a binary decision rule, there exist many equivalent systems that utilize POVMs with more than 2 outcomes. Briefly, the idea is that instead of summing together many lower-ranked operators to arrive at two operators that each correspond to a separate decision, we can equivalently perform a POVM with the lower-ranked operators and then use decision regions to sum the probabilities of their outcomes into two groups, each corresponding to a separate decision. Let us start with a system comprised of a 2-outcome POVM with elements $\{\tilde{E_0},\tilde{E_1}\}$, followed by a binary decision rule with decision regions $\tilde{\mathcal{D}} = \{1\}$ and $\tilde{\mathcal{D}}^C = \{0\}$. We would like to find the conditions under which a second, arbitrary system comprised of an $M$-outcome POVM with elements $\{E_m, 0 \leq m \leq M-1\}$, followed by a binary decision rule with decision regions $\mathcal{D} \subset \{0, 1, \dots, M-1\}$ and $\mathcal{D}^C$, is equivalent to the first. We can do so by using Equation \ref{eq:pm} to compute the probability that the second system decides `1' given that the input is $\rho_j$ for $j \in \{0,1\}$,
\begin{equation}
	\mathds{P}(\text{`1'} | \rho = \rho_j) = \sum_{m \in \mathcal{D}} Tr[E_m \rho_j] = Tr\left[\left(\sum_{m \in \mathcal{D}} E_m\right) \rho_j\right]
\end{equation}
where we have used the linearity of the trace operator. On the other hand, the probability that the first system decides `1' given that the input is $\rho_j$ is equal to $Tr[\tilde{E}_1\rho_j]$. Thus, in order for the two systems to be equivalent we must have
\begin{equation} \label{eq:povm_sum_over_decision_regions}
	\tilde{E_1} = \sum_{m \in \mathcal{D}} E_m, \quad \tilde{E_0} = \sum_{m \in \mathcal{D}^C} E_m.
\end{equation}
The second equality in Equation \ref{eq:povm_sum_over_decision_regions} follows from the first since $\tilde{E}_0 + \tilde{E}_1 = \mathds{1}_\mathcal{H}$ and $\sum_{m \in \mathcal{D}} E_m + \sum_{m \in \mathcal{D}^C} E_m = \mathds{1}_\mathcal{H}$. Thus, any $M$-outcome POVM and decision regions that satisfy Equations \ref{eq:povm_sum_over_decision_regions} will form a binary hypothesis testing system that is equivalent to the original system.

In summary, comparing Equations \ref{eq:benedetto_and_kebo} and \ref{eq:povm_sum_over_decision_regions} leads to the observation that the following systems are equivalent: (1) The system whose pre-decision operator is the POVM with elements $\{\tilde{E}_0, \tilde{E}_1\}$ and whose binary decision rule has decision regions $\tilde{\mathcal{D}} = \{1\}$ and $\tilde{\mathcal{D}}^C = \{0\}$, and (2) the system whose pre-decision operator is the POVM with elements $\{E_{mk} = \ket{\tilde{v}_{mk}}\bra{\tilde{v}_{mk}}\}$ and whose binary decision rule has decision regions $\mathcal{D} = \{\text{\emph{outcomes associated with the }} E_{1k}\}$, $\mathcal{D}^C = \{\text{\emph{outcomes associated}}$ $\text{\emph{with the }} E_{0k}\}$.

\subsection{Correspondence between Parseval Frames and Standard Measurements} \label{ssec:standard_measurements}

Our objective is to show that the elements $\{E_m, \, 0 \leq m \leq M-1\}$ of an arbitrary standard POVM can each be expressed as a sum of the form
\begin{equation} \label{eq:general_povm_frame_decomposition}
	E_m = \sum_{k=0}^{K_m-1} \ket{f_{mk}}\bra{f_{mk}}, \quad K_m \geq \text{rank}(E_m)
\end{equation}
for any integers $\{K_m\}$ with $K_m \geq \text{rank}(E_m)$, where the vectors $\{\ket{f_{mk}}\}$ are not unique. Since the $\{E_m\}$ satisfy the completeness relation $\sum_{m=0}^{M-1} E_m = \mathds{1}_\mathcal{H}$, Equation \ref{eq:general_povm_frame_decomposition} implies that the collection of vectors $\{\ket{f_{mk}}, \, 0 \leq m \leq M-1, \, 0 \leq k \leq K_m-1\}$ forms a Parseval frame for $\mathcal{H}$. The result in \cite{benedetto2008role} shows that Equation \ref{eq:general_povm_frame_decomposition} is satisfied for $K_m = \text{rank}(E_m)$ and $\ket{f_{mk}} = \ket{\tilde{v}_{mk}}$.

Consider a 2-outcome standard measurement with elements $E_1 = \mathcal{P}_\mathcal{V}$, $E_0 = \mathcal{P}_{\mathcal{V}^\perp}$, where $\mathcal{V} \subset \mathcal{H}$ is some subspace of $\mathcal{H}$ and $\mathcal{V}^\perp$ is its orthogonal complement. Note that rank($E_1$) = dim($\mathcal{V}$) and rank($E_0$) = dim($\mathcal{V}^\perp$). Let us assume that this measurement represents the pre-decision operator of a binary hypothesis testing system, and that it is followed by a binary decision rule with decision regions $\mathcal{D} = \{1\}$, $\mathcal{D}^C = \{0\}$. Both measurement elements can be written in many different ways as sums of rank-one operators,
\begin{subequations} \label{eq:project_onto_V}
\begin{alignat}{3}
	E_1 &= \mathcal{P}_\mathcal{V} &= \sum_{k=0}^{K_1-1} \ket{f_{1k}}\bra{\hat{f}_{1k}}, \quad &K_1 \geq \text{dim}(\mathcal{V}) \\
	E_0 &= \mathcal{P}_{\mathcal{V}^\perp} &= \sum_{k^\prime=0}^{K_0-1} \ket{f_{0k^\prime}}\bra{\hat{f}_{0k^\prime}}, \quad &K_0 \geq \text{dim}(\mathcal{V}^\perp).
\end{alignat}
\end{subequations}
Equations \ref{eq:project_onto_V} corresponds to decomposing any vector in $\mathcal{V}$ or $\mathcal{V}^\perp$ as a sum of its components in the directions defined by the $\{\ket{f_{0k^\prime}}\}$ or $\{\ket{f_{1k}}\}$, respectively. In particular, the $\{\ket{f_{1k}}\}$ and $\{\ket{\hat{f}_{1k}}\}$ can be any frame for $\mathcal{V}$ along with its corresponding dual frame. If $K_1 > \text{dim}(\mathcal{V})$ then the $\{\ket{f_{1k}}\}$ constitute an overcomplete representation for $\mathcal{V}$, whereas if $K_1 = \text{dim}(\mathcal{V})$ they are a basis for $\mathcal{V}$ and are complete but not overcomplete. Similar statements are true for the $\{\ket{f_{0k^\prime}}\}$ and $\{\ket{\hat{f}_{0k^\prime}}\}$. We wish to be able to interpret each rank-one operator $\ket{f_{1k}}\bra{\hat{f}_{1k}}$ or $\ket{f_{0k^\prime}}\bra{\hat{f}_{0k^\prime}}$ as a valid POVM element itself. Then using Equation \ref{eq:povm_sum_over_decision_regions}, we will be able to identify a rank-one POVM and a corresponding pair of decision regions that form an equivalent system to the one described above. According to the definition of a POVM as stated in Section \ref{ssec:postulates}, this means that each rank-one operator must be Hermitian and non-negative. Each condition individually implies that the components must have the form $\ket{f}\bra{f}$. In other words, the basis or frame must be self-dual. This means that only Parseval frames, of which orthonormal bases are of course a special case, can be used.

In summary, if $E_1 = \mathcal{P}_\mathcal{V}$ and $E_0 = \mathcal{P}_{\mathcal{V}^\perp}$ are written in the form of Equations \ref{eq:project_onto_V}
for two Parseval frames $\{\ket{f_{1k}}\}$  and $\{\ket{f_{0k^\prime}}\}$ for $\mathcal{V}$ and $\mathcal{V}^\perp$, respectively, then we will have $\ket{\hat{f}_{1k}} = \ket{f_{1k}}$ and $\ket{\hat{f}_{0k^\prime}} = \ket{f_{0k^\prime}}$. This will imply that each term in the two sums can themselves be elements of a separate POVM, so that the following two systems are equivalent: (1) the binary hypothesis testing system whose pre-decision operator is the 2-outcome POVM with elements $\{E_1, E_0\}$, and whose binary decision rule has decision regions $\mathcal{D} = \{1\}$ and $\mathcal{D}^C = \{0\}$, and (2) the binary hypothesis testing system whose pre-decision operator is the rank-one POVM with elements $\{E_{1k} = \ket{f_{1k}}\bra{f_{1k}}, 0 \leq k \leq K_1-1\}$ in addition to $\{E_{0k^\prime} = \ket{f_{0k^\prime}}\bra{f_{0k^\prime}}, 0 \leq k^\prime \leq K_0-1\}$ , and whose binary decision rule has decision regions $\mathcal{D} = \{\text{\emph{outcomes associated with the }} E_{1k}\}$ and $\mathcal{D}^C = \{\text{\emph{outcomes associated}}$ $\text{\emph{with the }} E_{0k}\}$. If either the $\{\ket{f_{0k^\prime}}\}$ or the $\{\ket{f_{1k}}\}$ form an overcomplete representation of $\mathcal{V}$ or $\mathcal{V}^\perp$, respectively, then the rank-one measurement is by definition non-standard. We may also sum multiple rank-one operators together to create non-rank-one POVMs, and the decision regions can be adjusted appropriately to create other equivalent systems.

When this result is applied to Helstrom's 2-outcome standard POVM in Equations \ref{eq:helstrom_subspace_criterion}, it follows that any two Parseval frames of $\mathcal{W}_1$ and $\mathcal{W}_0$ can be used to decompose $\Pi_1 = \mathcal{P}_{\mathcal{W}_1}$ and $\Pi_0 = \mathcal{P}_{\mathcal{W}_0}$ as sums of rank-one operators. The operators can themselves be interpreted as the elements of a rank-one POVM, which when followed by a binary decision rule whose decision regions group the outcomes into those associated with $\mathcal{W}_1$ and $\mathcal{W}_0$, achieves minimum probability of error. The orthogonal basis vectors $\{\ket{z_m}\}$ are one option for doing so, but any other Parseval frame is equally valid. This includes overcomplete representations of $\mathcal{W}_1$ and $\mathcal{W}_0$ that will then lead to non-standard measurements.%An interesting question is whether in this or other scenarios, overcomplete representations could be used as a way of improving the robustness of quantum binary hypothesis testing systems.

\subsection{Correspondence between Parseval Frames and Non-Standard Measurements} \label{ssec:non_standard_measurements}

It turns out that the same result is true for an arbitrary (potentially non-standard) POVM. Roughly, the idea is that any non-negative Hermitian operator $E$ acting on $\mathcal{H}$ can be written in terms of its real, non-negative eigenvalues and orthonormal eigenvectors as
\begin{equation} \label{eq:E_eigendecomposition}
	E = \sum_{i=0}^{N-1} \lambda_i \ket{v_i}\bra{v_i}.
\end{equation}
To rewrite $E$ as a sum of outer products of the form $\ket{f_k}\bra{f_k}$ for $0 \leq k \leq K-1$, we must find a Parseval frame for the range of $E$ and then rescale the elements of each frame vector in order to incorporate the $\{\lambda_i\}$.

Mathematically, we assume that for an arbitrary POVM the eigenvectors and eigenvalues of each element $E_m$ are known. The objective is to express each $E_m$ as a sum of outer products of the form $\ket{f_{mk}}\bra{f_{mk}}$ for $0 \leq k \leq K_m-1$, where $K_m$ is an integer of our choosing that must be at least as large as the rank of $E_m$. The problem can be solved independently for each value of $m$, so in what follows we will omit the subscript $m$ for ease of notation. Given an eigendecomposition of the form in Equation \ref{eq:E_eigendecomposition} in addition to an integer $K \geq \text{rank}(E)$, the objective is thus to find vectors $\{\ket{f_k}\}$ satisfying
\begin{equation} \label{eq:E_equals_sum_of_fkfk}
	E = \sum_{k=0}^{K-1} \ket{f_k}\bra{f_k}.
\end{equation}
We start by writing each $\ket{f_k}$ as a linear combination of the $\{\ket{v_i}\}$, $\ket{f_k} = \sum_{i=0}^{N-1} c_{ki} \ket{v_i}$. Substituting into the right-hand side of Equation \ref{eq:E_equals_sum_of_fkfk} and setting the result equal to the right-hand side of Equation \ref{eq:E_eigendecomposition} leads to the matrix equation
\begin{equation} \label{eq:coeff_matrix}
	A \, A^\dagger = \Lambda
\end{equation}
where $A$ is the $L \times K$ matrix, $L = max\{N,K\}$, whose $k$th column contains the coefficients $c_{k0} \dots c_{k,N-1}$ of the frame vector $\ket{f_k}$ in the $\{\ket{v_i}\}$ basis and $\Lambda$ is the $L \times L$ diagonal matrix with entries equal to the $\{\lambda_i\}$. Equation \ref{eq:coeff_matrix} says that the rows of $A$, which are vectors in $\mathds{C}^L$, are orthogonal and have squared norms $\lambda_0 \dots \lambda_{N-1}$. This allows us to conclude that the $\{\ket{f_k}\}$ can be constructed in the following way:
\begin{enumerate}
    \item Let $L = max\{N,K\}$. Start with any orthonormal basis of $\mathds{C}^L$. 
    \item Choose $N$ of the basis vectors and scale them so that they have squared norms $\lambda_0 \dots \lambda_{N-1}$. Let these vectors be the rows of the matrix $A$, and let the elements of the $k$th row be denoted by $c_{0k} \dots c_{K-1,k}$. Note that if the $i$th eigenvalue of $E$ is equal to zero, the $i$th row of $A$ will also be zero.
    \item For each $0 \leq k \leq K-1$, compute the frame vector $\ket{f_k}$ by weighting the vectors $\{\ket{v_i}\}$ by the elements of the $k$th column of $A$ and summing them together, $\ket{f_k} = \sum_{i=0}^{N-1} c_{ki} \ket{v_i}$. Then we will have $E = \sum_{k=0}^{K-1} \ket{f_k}\bra{f_k}$, as desired. Since we can start with any orthonormal basis of $\mathds{C}^L$, the $\{\ket{f_k}\}$ are not unique.
\end{enumerate}
If the above process is repeated for every $E_m$, then the collection of vectors $\{\ket{f_{mk}}, 0 \leq m \leq M-1, 0 \leq k \leq K_m-1\}$ will form a Parseval frame for $\mathcal{H}$. The result in \cite{eldar2002optimal} corresponds to choosing $K=1$ and choosing the initial orthonormal basis vectors in Step 1 to be the canonical basis vectors $\{\ket{g_\ell}\}$ of $\mathds{C}^L$. The result in \cite{benedetto2008role} also corresponds to choosing the initial orthonormal basis vectors in Step 1 to be the $\{\ket{g_\ell}\}$, but with $K=N$ instead of $K=1$.

As another example, consider the case where $K=N$ and the initial orthonormal basis vectors are complex exponentials with period $N$. Mathematically, the $k$th basis vector is $\ket{w_k} = \sqrt{N^{-1}} \sum_{\ell=0}^{N-1} \exp[-j2\pi k \ell / N] \ket{g_\ell}$. After scaling each basis vector to have the correct squared norm and constructing the matrix $A$, we find that
\begin{equation}
	\ket{f_k} = \sum_{i=0}^{N-1} \sqrt{\frac{\lambda_i}{N}} e^{-j2\pi i k/N} \ket{v_i}.
\end{equation}
It is straightforward to verify that $\sum_{k=0}^{N-1} \ket{f_k}\bra{f_k} = E = \sum_{i=0}^{N-1} \lambda_i \ket{v_i}\bra{v_i}$, as desired.

\section{Conclusions} \label{sec:conclusions}

We have proposed a hypothesis testing framework that encompasses the typical formulations of the $m$-ary hypothesis testing problem in both the classical and quantum settings. Considering classical ROCs in the context of this framework led us to define two operating characteristics for the quantum case, which were termed decision operating characteristics (QDOCs) or measurement operator characteristics (QMOCs) depending on how they were generated. Examples were given of both QDOCs and QMOCs. We showed that QMOCs generated using all standard measurements for $N=2$ were ellipses in the $P_F$-$P_D$ plane, centered at the point $(1/2,1/2)$. The derivation also applies to certain cases where $N>2$. The correspondence between classical frame theory and the theory of POVMs as used for quantum measurement was briefly discussed and an explicit procedure was derived that allows the elements of any POVM to be written as many different sums of rank-one operators. The procedure naturally led to a method for starting with an arbitrary hypothesis testing system and generating many equivalent systems. In particular, this means that the method can be used to construct many different systems that achieve minimum probability of error when distinguishing between two fixed density operators. Some utilize standard measurements whereas others utilize non-standard measurements.

\appendix

\section{Proof that QMOCs Generated using Standard Measurements are Ellipses} \label{sec:ellipse}

We wish to show that any QMOC generated according to the method described in Example 2 of Section \ref{ssec:qmocs}, in which 2-outcome standard measurements are used to distinguish between arbitrary density matrices $\rho_0$ and $\rho_1$ with $N=2$, is a rotated ellipse centered at the point $(1/2,1/2)$. The derivation also applies to the case where $N>2$ and the eigenvectors of $\rho_0$ and $\rho_1$ with nonzero eigenvalues span a 1- or 2-dimensional subspace of $\mathcal{H}$, as long as the standard measurements used to generate the QMOC have the following properties: The first two elements of the measurement, $E_0$ and $E_1$, should be analogous to those defined by Equation \ref{eq:qmoc_f1_f0}, but with the additional requirement that $\ket{f_1}$ and $\ket{f_0}$ should lie in the subspace spanned by the nonzero eigenvectors of $\rho_0$ and $\rho_1$. The other measurement elements must therefore project onto subspaces of the orthogonal complement of that subspace. Again the final decision is `1' if the measurement outcome associated with $E_1$ occurs and `0' if the measurement outcome associated with $E_0$ occurs. The other possible outcomes have zero probability of occurring and can be associated with either final decision. Essentially, this reduces the problem to that of distinguishing between two mixed states with $N=2$.

The coordinates of the QMOC in terms of the angle $\theta$ are given in Equations \ref{eq:qmoc_coordinates}. We start by subtracting $1/2$ from each coordinate. Then after making the substitutions $x = P_F-1/2$, $y = P_D-1/2$, $a = (a_0-a_1)/2$, and $b = (b_0-b_1)/2$, the coordinates become $x = a \cos \theta$, $y = b \cos (\theta-\alpha)$. To show that these coordinates represent a rotated ellipse centered at the origin, it is enough to show that they can be written in the form $x = q \cos \beta \cos t - r \sin \beta \sin t$, $y = q \sin \beta \cos t + r \cos \beta \sin t$ for some semi-major axis $q$, semi-minor axis $r$, angle of rotation $\beta$ from the horizontal, and parameter $t$ (which will prove inconsequential for our purposes). This can be done by finding the points on each curve with maximum $x$- and $y$-values and then setting their coordinates equal to one another. This leads to the following expressions for $q$, $r$, and $\beta$ in terms of $a$, $b$, and $\alpha$:
\begin{subequations} \label{eq:ellipse_parameters}
\begin{alignat}{1}
	\beta &= \frac{1}{2}\tan^{-1}\left(\frac{2ab\cos\alpha}{a^2-b^2}\right) \\[7pt]
	q &= \left[\frac{1}{2}\left(a^2+b^2+\frac{a^2-b^2}{\cos(2\beta)}\right)\right]^{1/2} \\[7pt]
	r &= \left[\frac{1}{2}\left(a^2+b^2-\frac{a^2-b^2}{\cos(2\beta)}\right)\right]^{1/2}.
\end{alignat}
\end{subequations}
It can be verified that when $\beta$, $q$, and $r$ are given by Equations \ref{eq:ellipse_parameters}, the coordinates $x = a \cos \theta$, $y = b \cos (\theta-\alpha)$ satisfy the equation that defines a (possibly rotated) ellipse centered at the origin: $\mathcal{A}x^2 + \mathcal{B}xy + \mathcal{C}y^2 + \mathcal{D} = 0$ with $\mathcal{B}^2-4\mathcal{A}\mathcal{C} < 0$, where $\mathcal{A}$, $\mathcal{B}$, $\mathcal{C}$, and $\mathcal{D}$ are functions of $q$, $r$, and $\beta$. This verifies the fact that the original QMOC is a rotated ellipse centered at the point $(1/2,1/2)$.

\bibliographystyle{IEEEbib}
\bibliography{refs}

\end{document}